\documentclass[aps,pre,onecolumn,superscriptaddress]{revtex4}
\usepackage[dvips]{graphicx}
\usepackage[dvips]{color}
\usepackage{breakurl,amsmath,amssymb,pst-node,eepic}

\usepackage{subfigure}
\usepackage{tikz}
\usetikzlibrary{shapes, arrows, calc, positioning,matrix}
\usepackage{bbold}

\begin{document}

\title{Reciprocity and success in academic careers}

\author{Weihua Li}
\email{weihuali89@gmail.com}
\affiliation{Department of Computer Science, University College London, 66-72 Gower Street, London WC1E 6EA, UK}
\affiliation{Systemic Risk Centre, London School of Economics and Political Sciences, Houghton Street, London WC2A 2AE, UK}

\author{Tomaso Aste}
\affiliation{Department of Computer Science, University College London, 66-72 Gower Street, London WC1E 6EA, UK}
\affiliation{Systemic Risk Centre, London School of Economics and Political Sciences, Houghton Street, London WC2A 2AE, UK}

\author{Fabio Caccioli}
\affiliation{Department of Computer Science, University College London, 66-72 Gower Street, London WC1E 6EA, UK}
\affiliation{Systemic Risk Centre, London School of Economics and Political Sciences, Houghton Street, London WC2A 2AE, UK}
\affiliation{London Mathematical Laboratory, 8 Margravine Gardens, London WC 8RH, UK}

\author{Giacomo Livan}
\email{g.livan@ucl.ac.uk}
\affiliation{Department of Computer Science, University College London, 66-72 Gower Street, London WC1E 6EA, UK}
\affiliation{Systemic Risk Centre, London School of Economics and Political Sciences, Houghton Street, London WC2A 2AE, UK}


\begin{abstract}
The growing importance of citation-based bibliometric indicators in shaping the prospects of academic careers incentivizes scientists to boost the numbers of citations they receive. Whereas the exploitation of self-citations has been extensively documented, the impact of re- ciprocated citations has not yet been studied. We study reciprocity in a citation network of authors, and compare it with the average reciprocity computed in an ensemble of null network models. We show that obtaining citations through reciprocity correlates negatively with a successful career in the long term. Nevertheless, at the aggregate level we show evidence of a steady increase in reciprocity over the years, largely fuelled by the exchange of citations between coauthors. Our results characterize the structure of author networks in a time of increasing emphasis on citation-based indicators, and we discuss their implzications towards a fairer assessment of academic impact.
\end{abstract}

\maketitle

The majority of measures that evaluate academic impact are based on citations of publications.
Ranging from crude citation counts to the well known journal impact factor \cite{garfield2006history}, the Eigenfactor metrics for journal rankings \cite{bergstrom2008eigenfactor} and the $h$-index \cite{hirsch2005index} for authors, such measures are increasingly relied upon to inform all aspects of academic decision-making, including faculty recruiting, grant attribution, and the formation of collaborations \cite{clauset2017data}. The onset of such a trend dates back to the 1950s, when the earliest citation-based indices to assess academic impact were put forward \cite{garfield1955citation}. The following decades saw the proliferation and ever-increasing adoption of such indicators in all fields of science \cite{garfield1970citation, bornmann2008citation}, which eventually led to the systematic analysis of academic citations, to the emergence of \emph{bibliometrics} as a research field \cite{hood2001literature}, and, more recently, to the rise of a ``science of science'' devoted to understanding the determinants of scientific success \cite{petersen2014reputation, weiss2014adoption, fortunato2018science}.   

Given the importance that citations play nowadays in shaping the prospects of an academic career, it is certainly not surprising to see an increased attention to the study of citation patterns and of the publication strategies that can attract a larger number of citations. While an author's productivity and the quality of her work are obvious determinants of academic success \cite{sinatra2016quantifying}, other less tangible aspects have recently been identified as key contributors to success. For example, the social network a scientist is embedded in has recently been shown to play a relevant role in determining her future chances of success \cite{sarigol2014predicting}.

Other social and behavioural
considerations also play relevant roles in modern academic reputation systems, especially concerning the proliferation of self-citations and citations between close collaborators \cite{hellsten2007self, fowler2007does, radicchi2009diffusion, martin2013coauthorship}. Self-citations and their role in inflating bibliometric indicators have been studied extensively \cite{seeber2017self}. Several scholars have proposed revised metrics to mitigate their distorting impact \cite{bartneck2010detecting}, and indeed some of the most popular citation indexing services (e.g., Web of Science) and social networking sites for scientists (e.g., ResearchGate) provide detailed author-specific bibliometric information with and without the contribution of self-citations.

Yet, very little attention has been paid to the 
{scientific community's collective response to the increasing adoption of bibliometric indicators, and whether this ultimately resulted in more sophisticated citation patterns involving collaborators and colleagues \cite{guimera2005team, martin2013coauthorship}}. Indeed, the quest for higher citation counts generates an obvious incentive for scientists to exchange citations with their closest circle of coauthors and collaborators. Such behaviour, known as reciprocity, has long been observed and investigated in many social and economic systems \cite{trivers1971evolution,berg1995trust, hilbe2018partners}. It reflects the tendency to return helpful acts, and it is crucial for forming and maintaining cooperative relationships among individuals and groups \cite{mahmoodi2018reciprocity}.

Following the broad stream of literature that has analyzed citation patterns from a network perspective (see, e.g., \cite{newman2004coauthorship}), we investigate the inherent reciprocity dynamics in an author citation network constructed from the citation history of $463,348$ papers in the \emph{Physical Review} (PR) corpus of journals published by the American Physical Society (APS) between 1893 and 2010 (see Appendix \ref{sec:methods}), spanning all research fields of physical sciences, with disambiguated author names obtained from \cite{sinatra2016quantifying}.

We address two main research questions: $1$) how much reciprocated citations contribute to a scientist's academic reputation, and whether these can be used to predict success and classify different career trajectories; $2$) how prevalent is the tendency to reciprocate citations in the scientific community at large, and how it has evolved over time. 

We tackle such questions by representing authors as nodes in a directed weighted network $\mathcal{C}$, where the weight $c_{ij}$ denotes the number of times author $i$ has cited author $j$, and use such a representation to measure the number of reciprocated citations as $c^\leftrightarrow_{ij} = \min(c_{ij},c_{ji})$. We assess the statistical significance of the empirically observed patterns of reciprocated citations with the values observed under an ensemble of null network models obtained through the controlled randomization of the original networks' topology (see Appendix \ref{sec:methods}).

Following the definition in \cite{garlaschelli2004patterns}, we define the \emph{excess} reciprocity
\footnote{The quantity in Eq. \ref{recp} was simply referred to as reciprocity in \cite{garlaschelli2004patterns}
, where it was first introduced. We shall instead refer to it as excess reciprocity in order to highlight that such a definition always entails a comparison with a benchmark defined by a null model} of an author $i$ as
\begin{equation}
	\rho^{(i)} = \frac{\rho_0^{(i)} - \langle \rho_{\mathrm{null}}^{(i)} \rangle}{1 - \langle \rho_{\mathrm{null}}^{(i)} \rangle} \ ,
	\label{recp}
\end{equation}
where $\rho_0^{(i)} = \sum_{j \in \mathcal{C}} c^\leftrightarrow_{ij} / \sum_{j \in \mathcal{C}} c_{ji}$ is the fraction of citations received by author $i$ in the empirical author citation network which she reciprocated (i.e., the fraction of reciprocated incoming weight), while $\langle \rho_{\mathrm{null}}^{(i)} \rangle$ is the average of the same quantity as computed in the null model ensemble. As we shall explain more extensively later on, the rationale of the above definition lies in its ability to quantify an \emph{author-specific} propensity to reciprocate with respect to an expected benchmark quantified by the null model. For example, an excess reciprocity $\rho^{(i)} = 0.1$ for an author with $100$ received citations would correspond to $55$ of them being reciprocated under an expected reciprocity $\langle \rho_{\mathrm{null}}^{(i)} \rangle = 1/2$, and to $28$ reciprocated citations under an expected reciprocity $\langle \rho_{\mathrm{null}}^{(i)} \rangle = 1/5$. 

\section{Results}

\subsection{Reciprocity and career success.}

\begin{figure*}[h!]
	\centering
	\includegraphics[width=.6\linewidth]{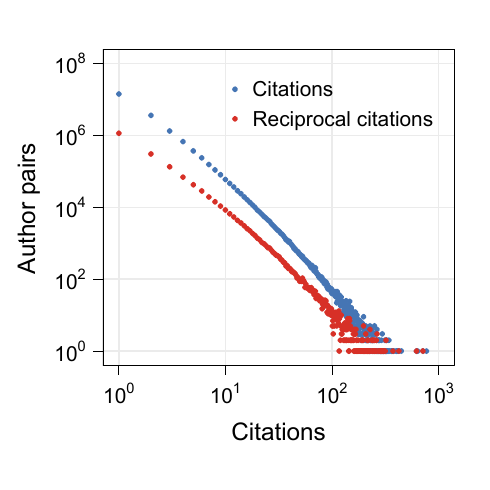}
	\caption{Frequency of citations between pairs of authors. Empirical frequency of the number of directed (blue dots) and reciprocated (red dots) citations between pairs of authors active in the APS dataset between 1950 and 2010.}\label{fig:pair}
\end{figure*}

Fig. \ref{fig:pair} shows the frequency of both the number of directed citations (i.e., the number of times $c_{ij}$ an author $i$ has cited another author $j$) and the number of reciprocated citations (i.e., $c_{ij}^\leftrightarrow = \min (c_{ij},c_{ji})$) across all pairs of authors active in the APS dataset from $1950$ to $2010$. As it can be seen, both distributions show a markedly heavy-tailed behaviour. In particular, there are more than fifty thousand pairs of authors with $10$ reciprocated citations, and more than two thousand pairs with $50$ or more reciprocated citations. {Overall, more than $21\%$ of the citations in the dataset are reciprocated (roughly $8.5$ millions out of $40.4$ millions).}

\begin{figure*}[h!]
\vspace{-1.5cm}
\centering
\includegraphics[width=.96\linewidth]{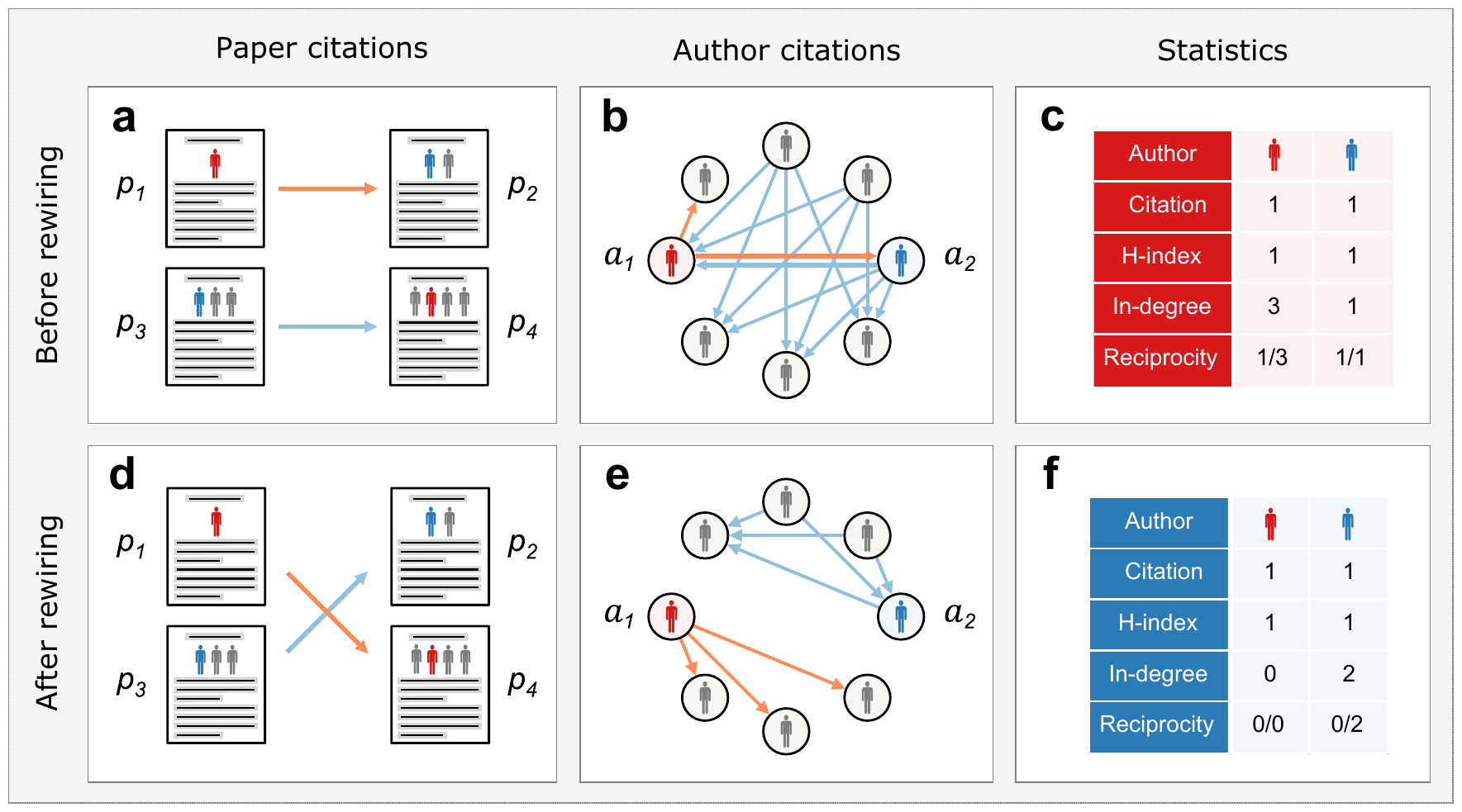}
\caption{Illustration of the geodesic distance $1$ null network model. Randomly selected pairs of links representing citations between papers are swapped with probability $1/2$ whenever they fulfil two conditions on time and ``distance'' (see Methods). Panels in the Figure illustrate the main ingredients and steps of the null model.
		\textbf{a} Two randomly selected pairs of citations $p_1 \rightarrow p_2$ and $p_3 \rightarrow p_4$ between papers $p_j$ $(1 \leq j \leq 4)$. We highlight two authors $a_1$ and $a_2$ in orange and blue respectively, to draw attention to the aforementioned ``distance'' constraint, i.e., that links are rewired only when the papers they connect either share at least one common author, or there exists at least one citation between authors from the two papers in the author citation network (see Methods).
		\textbf{b} The citation network of authors constructed from the papers in \textbf{a}, where we highlight authors $a_1$ and $a_2$ and the presence of a pair of reciprocated citations between them.
		\textbf{c} Citation-based indicators and statistics of authors in the network. 
		\textbf{d} Rewiring of the links, which leads to two new citations $p_1 \rightarrow p_4$ and $p_2 \rightarrow p_3$.
		\textbf{e} The citation network of authors constructed from \textbf{d} after removing self-citations.
		\textbf{f} Citation-based indicators and statistics of authors after the rewiring. As it can be seen, the number of citations received and the $h$-index of all authors are preserved. }\label{fig:illustrate}
\end{figure*}

The sheer scale of the aforementioned phenomenon suggests that reciprocated citations might play a key role in shaping an author's scientific success. Does reciprocity pay? Do authors with a history of systematic reciprocity outperform their peers? In this section we investigate citation patterns at the level of individual authors in order to answer these questions. We do so by proposing a null network model ensemble (see Fig. \ref{fig:illustrate} and Methods) to estimate the average baseline level of reciprocity $\langle \rho_{\mathrm{null}}^{(i)} \rangle$ one should expect in the author citation networks under partially random interactions, which we then use to measure excess reciprocity.

The rationale of the definition in \eqref{recp} is to discount density-related effects. Indeed, simply measuring reciprocity as the fraction of reciprocated weight $\rho_0^{(i)}$ typically leads to seemingly high (low) values in dense (sparse) networks. The measure in Eq. \eqref{recp} takes care of such potential spurious effects by discounting the average reciprocity observed in a null model ensemble, so that positive (negative) values of $\rho^{(i)}$ indicate authors whose citations have been received through an over-representation (under-representation) of reciprocated relationships, whereas values $\rho^{(i)} \simeq 0$ indicate levels of reciprocity compatible with the null assumption encoded in the network null model being used. In conclusion, excess reciprocity indirectly quantifies how much the academic reputation of an author, as measured by her number of citations and $h$-index (which are both preserved by the null model, see Methods), relies on the citations from authors she cited as well.

We investigated the relationship between excess reciprocity and success by following the career paths of authors with a traceable publication history of at least $20$ years in the APS dataset. We first employed a variety of methodologies to predict a scientist's future success (in terms of citations\footnote{Let us remark here that it is not our intention to equate academic success, which is multifaceted, to mere citation counts. However, we take the position that a high citation count is an unquestionable indicator of success for a scientist's production.}) based on her previous history of excess reciprocity. In all cases we found very weak to no evidence of any predictive power (see Appendix \ref{sec:regression}), which strongly suggests that citation strategies based on the mere exchange of citations do not contribute to attracting higher numbers of citations in the future.  

We then applied the $k$-means clustering algorithm \cite{macqueen1967some} to categorize authors in terms of career success. Following \cite{sinatra2016quantifying}, we performed this analysis considering the career trajectories of all authors with $10$ or more papers published over the course of at least $20$ years (with at least one paper published every $5$ years) who published their first paper either in $1950$-$1970$ or in $1970$-$1990$. We used these two samples to perform a $k$-means clustering analysis based on the cumulative number of citations received over time. Since several authors did not receive citations early in their career, we performed our analysis starting from the $4$th career year.

In Fig. \ref{fig:micro} we present the results for $1970$-$1990$ (see Appendix \ref{sec:ind_excess} for the results obtained for $1950$-$1970$), which were obtained on a pool of $5,070$ scientists. We identified $4$ distinct groups with very different levels of career success, ranging from a small minority of authors ($1.2\%$ of the sample) who managed to attract several thousand citations over the time period considered, to the relative majority of authors ($67.4\%$ of the sample) who only enjoyed moderate to low success (see Fig. \ref{fig:micro} caption for more details).

\begin{figure}[h!]
	\centering
	\includegraphics[width=.96\linewidth]{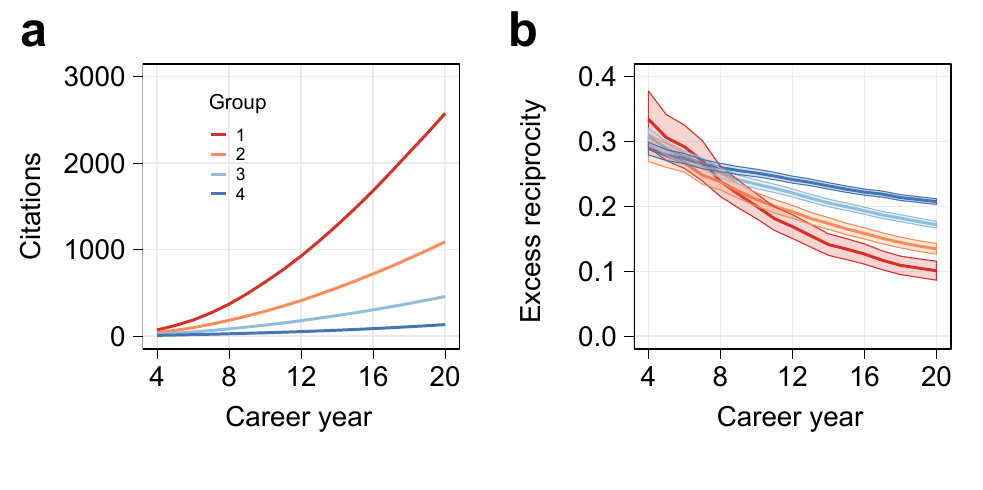}
	\caption{Illustration of the relationship between excess reciprocity and long-term career success.
		\textbf{a} Centroids of the clusters identified by $k$-means based on the cumulative number of citations received by authors who started their career between $1970$ and $1990$. The sample contains $5,070$ authors, and the fraction of authors falling within each cluster are as follows: $1.2\%$ in group $1$, $6.9\%$ in group $2$, $24.5\%$ in group $3$, $67.4\%$ in group $4$. 
		\textbf{b} Excess reciprocity, defined as per Eq. \eqref{recp}, within each cluster. Thick solid lines denote the average within the cluster, while ribbon bands denote $95\%$ confidence level intervals. 
	}\label{fig:micro}
\end{figure}

We find the above groups to be associated with markedly different behaviours. Namely, we find long-term career success to be associated with lower levels of excess reciprocity. Indeed, the two most successful groups are associated with the lowest long-run excess reciprocity, with the small cluster of elite scientists (group $1$ in Fig. \ref{fig:micro}) displaying an average excess reciprocity around $0.1$ towards year $20$ of their career. Conversely, the two least successful groups are associated with consistently higher levels of long-term excess reciprocity, higher than $0.2$ in the case of the single least successful group.

In addition, we checked whether authors belonging to a certain group tend to publish more frequently in some APS journals rather than others. The results of this analysis are presented in Appendix \ref{sec:ind_excess}, and show that authors in the most successful groups (group $1$ and $2$) have a higher publication rate in \emph{Physical Review Letters} (PRL), which is somewhat unsurprising since PRL is by far the most impactful venue among those considered here. Yet, a more nuanced picture emerges when looking at the remaining journals, as the most successful clusters do not necessarily account for the relative majority of publications in the most impactful journals and vice versa. Moreover, while the ranking and behaviours in terms of excess reciprocity are similar across the two time periods we consider, it is interesting to notice that publication rates of the different groups across journals are rather different (see Appendix \ref{sec:ind_excess}).  

We tested the robustness of the negative relationship between excess reciprocity and career success in a number of ways, in order to rule out  spurious effects due to possible confounding factors. First, following \cite{guimera2007classes} we modified our null model in order to account for modularity-related effects in the author network, i.e., that authors belonging to the same scientific sub-communities can be naturally expected to exchange citations at an above average rate. To this end, we used two popular community detection algorithms (InfoMap \cite{rosvall2008maps} and the modularity-based algorithm published in \cite{clauset2004finding}) in order to extract the community structure of the author network at different granularity levels (see Appendix \ref{sec:alt_null}), and constrained our null model to partially preserve it (see Methods).

Second, we proceeded to discount self-citation as a potential driver of reciprocity (e.g., coauthors both citing their own past work) by repeating our analyses after removing all self-citations in the paper network.

Third, we discounted the impact of authors with very low productivity and impact (at least within the APS corpus) by repeating our analyses after removing all authors with a total of less than $10$ citations over their first $20$ career years (see Methods).

In order to be as thorough as possible, we applied both of the aforementioned community detection methods to the full dataset of authors who started their careers in 1970-1990, to the data after the removal of all self-citations, and to the data after the removal of all authors with less than $10$ citations (results shown in Appendix \ref{sec:alt_null}). In all cases we found the same negative relationship between excess reciprocity and success based on the four clusters of authors identified with $k$-means. As a further robustness check, we tested such relationship when separating authors based on different clustering criteria, i.e., we also grouped authors based on quartiles and with the Expectation-Maximization clustering algorithm \cite{dempster1977maximum}. In both cases, we still detected the same negative relationship (see Appendix \ref{sec:alt_null}).

Lastly, we tested such relationship from the opposite perspective, i.e., by grouping authors based on excess reciprocity and then measuring the success of different groups. We resorted to matched pair analysis, and divided the authors whose careers started in 1970-1990 into ``high reciprocators'' (treatment) and ``low reciprocators'' (control) groups based on their excess reciprocity pattern over the first $10$ career years, and performed a $t$-test on the average number of citations attracted by authors in the two groups over the following $10$ career years after pairing them based on productivity (i.e., on the number of papers published in the first $10$ years). Consistently with our results based on clustering, we found the treatment group to attract substantially less citations per author ($272.2$) than the control group ($331.6$), with $p < 0.001$ (see Appendix \ref{sec:matched}).

\subsection{Shifts in citation patterns.}

In the previous section we analyzed the relationship between excess reciprocity and long-term success from a cross-sectional point of view by ``collapsing'' together the career trajectories of several authors whose actual careers developed asynchronously over the span of a few decades. We now seek to further unpack this relationship by investigating temporal snapshots of the APS citation network, testing how an author's propensity to reciprocate citations correlated with her impact during different historical periods.

We do so by performing analyses at the decade level. For each decade from the $1950$s to the $1990$s, we consider all authors whose career started before the end of such decade and did not end before the first year of that decade. We then pool all the paper published by such authors before the end of the decade, and their corresponding citations, to build the author citation network for the decade of interest.

\begin{figure*}[h!]
	\centering
	\includegraphics[width=.96\linewidth]{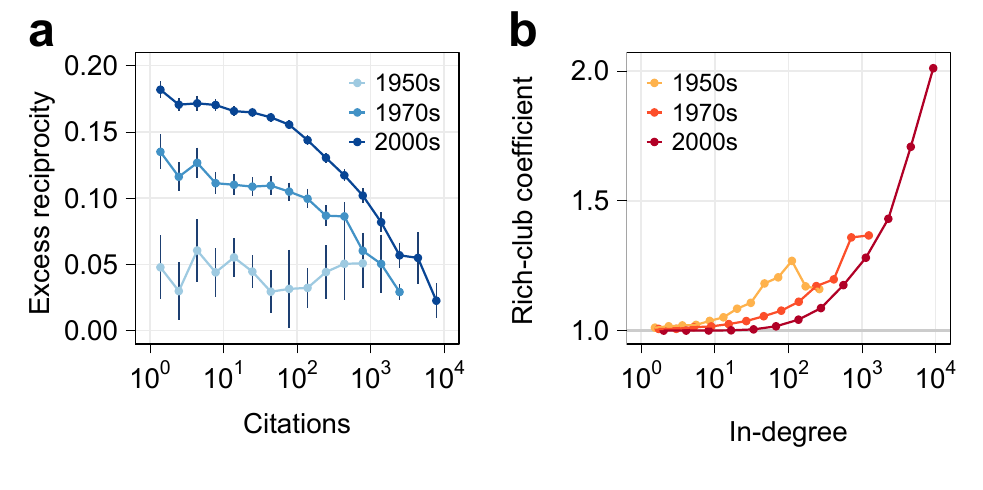}
	\caption{Results at the level of groups of authors.
		\textbf{a} Excess reciprocity as a function of the number of citations received. Error bars denote $95\%$ confidence level intervals.
		\textbf{b} Rich-club coefficient of the APS author citation network with respect to the null model (see Methods) in the $1950$s, $1970$s, and $2000$s. Here in-degree refers to the total weight received by the author in the author citation network.
	}\label{fig:meso}
\end{figure*}

Fig. \ref{fig:meso}a shows, for three different decades, the average excess reciprocity of authors as a function of their accrued citations (see Appendix \ref{sec:decade} for all six decades). As it can be seen, over time we observe the emergence of a clear negative correlation between an author's impact and her excess reciprocity. In the $1950$s the entire APS scientific community was essentially compatible with the null model, with average excess reciprocity lower than $0.05$ for all groups of authors. This changes considerably from the $1970$s onwards, and it becomes quite pronounced in the $2000$s, with a very apparent negative relationship between an author's impact and her tendency to reciprocate.  

One might intuitively expect high impact authors to display, as a group, the lowest tendency to reciprocate. Indeed, in network terms, highly successful academics simply do not have enough outgoing links to reciprocate their incoming links, i.e., they cannot provide enough citations to match the high number of citations they receive. While this is certainly true, as shown consistently for all decades in Fig. \ref{fig:meso}a, there are subtler aspects to this observation. 

First, let us recall that the definition in Eq. \eqref{recp} measures the excess of reciprocity with respect to an \emph{expected baseline}, which in our case is computed from a null model which preserves the heterogeneity (in terms of number of publications) and level of success (both in terms of accrued citations and $h$-index) of each author. In this respect, the above result shows that high impact authors simply do not reciprocate much more than one could reasonably expect. Yet, a deeper analysis of the citations received by high impact authors reveals more substantial differences with respect to our null model. Indeed, while our null model naturally incorporates the low levels of excess reciprocity of high impact authors, it does not prescribe \emph{who} the recipients of citations from them should be.

To investigate who the recipients are, we examine the level of interconnectedness among the leading authors with the highest citation counts in each decade by measuring the rich-club coefficient \cite{zhou2004rich,opsahl2008prominence} in the author citation networks (see Methods Section). The rich-club coefficient quantifies the tendency to preferentially establish relationships within a group with respect to the expected tendency based on a null hypothesis. In the present case, we measure the rich club coefficient as $\phi(c) = \phi_0(c) / \langle \phi_{\mathrm{null}}(c) \rangle$, where $\phi_0(c)$ is the fraction of the total number of citations exchanged between authors that have received at least $c$ citations in the empirical network (i.e., authors with an incoming weight equal to or larger than $c$), and $\langle \phi_{\mathrm{null}}(c) \rangle$ is the corresponding quantity computed over our null network model ensemble.

We observe an increasingly pronounced rich-club effect among leading academics, with the effect being up to twice as strong with respect to the null model for authors with an incoming weight around $10^4$ in the $2000$s (Fig. \ref{fig:meso}b). Conversely, in earlier decades we find the effect to be much less strong, although still present (see also Appendix \ref{sec:decade}). This result indicates that, although the overall tendency of high impact authors to reciprocate is close to the one predicted by our null model, they overwhelmingly tend to cite their peers. The presence of such an interconnected rich core of successful scientists suggests that homophily with respect to career excellence has increasingly become one of the driving forces behind the attribution of citations. 

\subsection{Reciprocity, coauthorship and self-citations.}

We now shift our attention to the evolution of reciprocity at the aggregate level of the entire APS community. We straightforwardly generalize Eq. \eqref{recp} to define a measure of network-wide excess reciprocity as $\rho = (\rho_0 - \langle \rho_\mathrm{null} \rangle) / (1 - \langle \rho_\mathrm{null} \rangle)$, where $\rho_0$ denotes the overall fraction of reciprocated weight in the empirical networks, whereas $\langle \rho_\mathrm{null} \rangle$ denotes the corresponding average quantity in the null network ensemble. We track such quantity over time by considering annual networks constructed by including all papers published by active authors up to the year under analysis. We consider an author to be active whenever the year under consideration is between the first and last of her career.

\begin{figure}[h!]
	\centering
	\vspace{-1.8cm}
	\includegraphics[width=.92\linewidth]{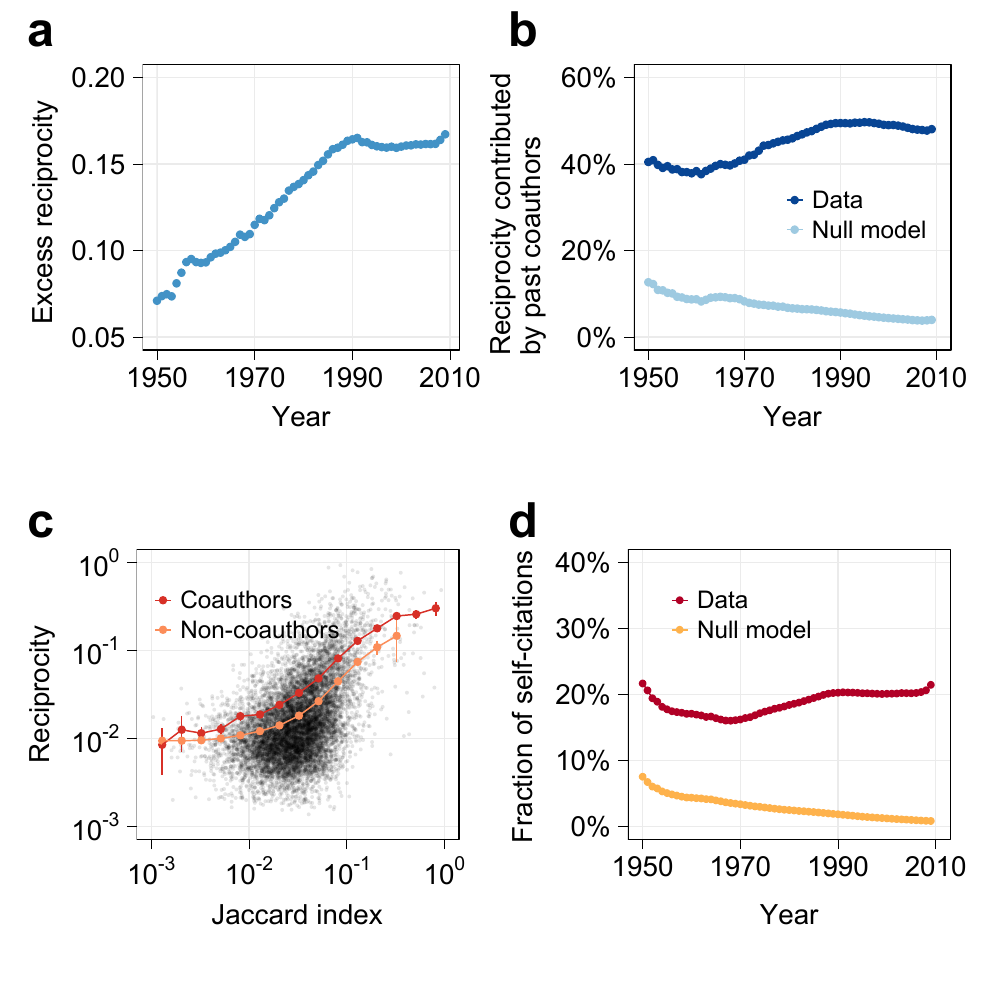}
	\vspace{-.8cm}
	\caption{Network-wide reciprocity in the empirical networks and in the null model.
		\textbf{a} Annual excess reciprocity values from 1950 to 2009.
		\textbf{b} Fraction of reciprocity $\rho_{0}$ contributed by past coauthors in the empirical networks (dark blue) and in the null model (light blue).	 
		\textbf{c} Fraction of reciprocated citations between two authors versus the overlap in their research interests, proxied by the Jaccard index between the list of references cited by the two authors, with $95\%$ confidence level intervals. The grey dots represent $10^4$ author pairs sampled from real data. When controlling for the overlap, on average coauthors are found to reciprocate more than non-coauthors.
		\textbf{d} Fraction of self-citations in the empirical networks (dark red) and in the null model (orange). The standard error bars for the null model results are small and not graphically visible in \textbf{b} and \textbf{d}.
	}\label{fig:macro}
\end{figure}

During the entire period of study we systematically observe positive values of network-wide excess reciprocity, indicating a stronger propensity of the APS community to exchange citations than the one expected in our null model ensemble.  Furthermore, we find reciprocity to increase steadily (and roughly linearly) up to the early $1990$s, after which it stabilizes around $0.15$ (Fig. \ref{fig:macro}a and Appendix \ref{sec:constraints}). 

A closer look reveals that, over the entire period of observation, a substantial proportion of the overall reciprocity $\rho_0$ is accounted for by citations between coauthors. Such proportion grows from about $40\%$ in the $1950$s to about $50\%$ in the $1990$s. This is in contrast with the expected proportion computed in the null model, which instead shows a steady decline over time (Fig. \ref{fig:macro}b). 

In order to better understand the impact of citations from coauthors on a scientist's career, we pool all authors over the entire period of observation and compare the tendency to reciprocate between coauthors and non-coauthors. 
Namely, we define the reciprocity between a \emph{pair} of authors $i$ and $j$ as the number of reciprocated citations between them, divided by the total number of citations received by both authors. In Fig. \ref{fig:macro}c we show such quantity as a function of the distance between research interests, quantified by the Jaccard similarity index between the sets of papers cited by a pair of authors over their career \cite{ciotti2016homophily}. Higher Jaccard indices indicate higher proportions of common references used by both authors, which we interpret as a proxy for a substantial overlap of research interests. As one would intuitively expect, we observe an overall positive correlation between research interests and the tendency to exchange citations. However, on average we find this relationship to be stronger in the case of coauthors, regardless of the specific level of proximity between research interests.

Let us conclude this section with a short digression devoted to the investigation of self-citations through the lens of our null model. Fig. \ref{fig:macro}d compares the observed fraction of self-citations with the corresponding expected proportion in our null model ensemble over time. As it can be seen, the empirical rate of self-citation has remained around a fairly stable level of around $20\%$ (which decreases to roughly $18\%$ when including authors no longer active in the time frame under consideration). Yet, the null model predicts a sharp downward trend, which, as in the case of reciprocity between coauthors, highlights a growing gap between empirical citation patterns and those expected under our null hypothesis. Interestingly, the aforementioned rate of self-citation is much larger than those observed in citation datasets from other disciplines (e.g., Law, Political Science, Mathematics), which in most cases are between $5\%$ and $12\%$ \cite{king2017men}.

\section{Discussion}

This study addressed two main research questions, i.e., whether the constantly increasing attention to citation counts and bibliometric indicators has led to an incentive for academics to boost such metrics through the exchange of citations, and, if so, whether this {behaviour} is rewarding in terms of career success. Let us stress from the outset that our results, being based on a comparison between empirical data and null network models, cannot provide direct evidence of strategic behaviour explictly aimed at manipulating bibliometric indicators. Yet, they inform us on how the academic community has collectively organized in response to the increased emphasis on such metrics, revealing a nuanced picture which we discuss in the following.

Starting from the $1950$s, we observe two main eras in the APS citation landscape. From $1950$ to $1990$ we observe a steady increase in network-wide excess reciprocity, which then remains roughly constant around $0.15$ for the following twenty years. It is tempting to relate the former trend with the onset and the ensuing rise of Scientometrics as a research field. Indeed, the concept of bibliometric indicators was first proposed in $1955$ by Garfield \cite{garfield1955citation}, and put in practice in $1964$ with the launch of the Science Citation Index, with an ensuing proliferation of indicators \cite{garfield1970citation} and the establishment of the first academic journal entirely devoted to Scientometrics in $1978$ \cite{de1978editorial}. 

It is somewhat surprising to observe the above trend plateauing over the last two decades, when the average number of citations made by papers in the APS increased \cite{martin2013coauthorship}, and the incorporation of citations and bibliometric indicators in academic decision-making witnessed further increase \cite{oecd2016compendium}. Yet, this is accompanied by a widening gap between the observed reciprocity taking place between coauthors (which, after the $1990$s accounts for almost \emph{half} of the whole network's reciprocity), and the expected level of reciprocity between coauthors in our null model. These two seemingly at odds trends might be related to the onset of the Internet, which dramatically reduced barriers to access to published research. We speculate that the expansion of the Internet might have, on average, widened a scientist's potential pool of authors to cite \cite{martin2013coauthorship}, therefore diluting the overall reciprocity in the scientific community while at the same time resulting in a more prominent role of the exchanged citations between collaborators and colleagues. 

The above change is accompanied by a structural evolution in the author-author citation network. Over the years we witness the emergence of a very clear rich club of top scientists who exchange citations at a rate much higher than the one predicted by our null model.
When looked at from the perspective of the whole network, however, the impact of such a rich club gets diluted, and top scientists end up being below average ``reciprocators''. Starting from the $1970$s the author network organizes in such a way that most of the reciprocity takes place at the level of low to medium impact authors. 

The latter result is corroborated by our clustering analysis (see Fig. \ref{fig:micro} and Appendix \ref{sec:alt_null}), which shows that only low-impact career trajectories tend to be associated with high excess reciprocity. All in all, these results suggest that even potentially sophisticated shortcuts to artificially boost bibliometric indicators based on the exchange of citations, rather than on mere self-citations, cannot circumvent the fact that consistently high-quality publications are, by far, the main determinant of academic success.

Yet, our results show that a remarkable proportion of most scientists' citations come, on average, from their immediate ``neighbourhood'' (Fig. \ref{fig:macro}b), and that the gap between such proportion and the one expected under our null hypothesis has been constantly increasing for more than $40$ years. We interpret this as an echo of the academic community's \emph{collective} incentive to boost individual metrics of academic reputation. In this respect, it is worth stressing {once more} that our results are of a statistical nature. As such, they cannot provide insight about the countless reasons that might induce \emph{individual} authors to reciprocate citations. A proper investigation of potentially malicious practices deliberately aimed at boosting bibliometric indicators is well beyond the scope of the present work, and most likely would entail collecting data by interviewing authors.

We believe our results should caution against the current practice of condensing the entirety of a scientist's production into a single bibliometric indicator, i.e., into a single number. We argue that, in analogy to what some academic platforms do to display information about a scientist's impact with and without self-citations, such a profiling should be complemented by providing a more detailed breakdown about the origin of a scientist's citations.

\appendix

\section{Methods}
\label{sec:methods}

\subsection{APS citation data and network construction.}

The APS corpus of journals publish articles spanning all research fields in Physics since 1893. 
The dataset contains all papers published between 1893 and 2010, which we complemented with the work done to disambiguate authors in \cite{sinatra2016quantifying}, where $236,884$ unique authors were identified. Following \cite{sinatra2016quantifying}, we only retained papers in the dataset with no more than 10 authors, which left us with $415,342$ papers and $4,125,843$ citations.

We used the above data to construct networks of citations both between papers and between authors.
{Although our analysis starts from $1950$, we still employed all papers published from the beginning of the APS dataset in order to account for the fact that reciprocity in citations is an inherently cumulative phenomenon.} Paper networks are directed and unweighted, while author networks are directed and weighted, with a link of weight $c_{ij}$ from node $i$ to node $j$ denoting that author $i$ has cited $c_{ij}$ times author $j$. We used this general structure to extract annual networks, by selecting all the active authors who had started their careers before and ended after the year of interest, and choosing all papers published by such authors before that same year.

In analogy, in order to perform our analyses at the decade level, we selected all authors active in at least one year of the decade of interest. We then retained all papers published by the selected authors before the end of the decade.

\subsection{Geodesic distance $1$ null network model.}

We propose a null network model ensemble to estimate an average baseline level of reciprocity $\langle \rho_\mathrm{null}^{(i)} \rangle$ one should expect to observe from a certain author $i$ in the citation networks under partially random interactions. This quantity can be then used to compute the author's excess reciprocity as per Eq. \eqref{recp}.

We build on and generalize the null network model put forward in \cite{livan2017excess} to measure reciprocity in directed unweighted networks. Given a set of papers, we construct our null network model ensemble according to the following principles. First, the ensemble should reflect the fact that citations are exchanged via papers, i.e., that papers are the fundamental units of interaction. Second, it should preserve the inherent directionality of citations between papers due to time ordering, i.e., that more recent papers cite older ones, and not the other way around. Third, the ensemble should account for the fact that most citations occur within well defined scientific communities, which can be proxied in terms of a ``distance'' constraint between authors, consistently with the vast amount of literature which shows that homophily and peer influence typically do not extend beyond a few degrees of separation in a variety of social networks \cite{christakis2007spread,li2017three,centola2010spread}.

Following the above prescriptions, we define our null network model ensemble as follows. We start from the real citation network structure, and iteratively select random pairs of citations $p_1 \rightarrow p_2$ and $p_3 \rightarrow p_4$ (see Fig. \ref{fig:illustrate}), where $p_i$ ($1 \leq i \leq 4$) denotes a paper and $p_i \rightarrow p_j$ denotes a citation from $p_i$ to $p_j$. With probability $1/2$ we swap the two links representing the citations if the two following conditions are both met: $i$) both new citations $p_1 \rightarrow p_4$ and $p_3 \rightarrow p_2$ preserve the time ordering of publication dates (i.e., $p_1$ was published before $p_4$, and $p_3$ was published before $p_2$), and $ii$) both new citations $p_i \rightarrow p_j$ (with $(i,j) \in \{(1, 4), (3, 2)\}$) are such that either papers $p_i$ and $p_j$ share at least one common author, or there exist authors $a_i$ from $p_i$ and $a_j$ from $p_j$ such that $a_i$ has cited $a_j$ or $a_j$ has cited $a_i$ at least once in the empirical citation network. The latter condition encodes the aforementioned distance constraint, limiting the set of allowed link rewiring moves to papers whose author lists feature at least one pair of authors at geodesic distance $1$ in the empirical author network (see Appendix \ref{sec:constraints} for the results obtained with a less constrained null model). When controlling for modularity-related effects, we retain the same rewiring procedure, but only accept rewiring moves when the new potential citations $p_i \rightarrow p_j$ (with $(i,j) \in \{(1, 4), (3, 2)\}$) are such that there exist authors $a_i$ from $p_i$ and $a_j$ from $p_j$ who both belong to the same community (see Appendix \ref{sec:alt_null}).

The above operations are repeated until the system has reached an equilibrium state, at which point we collect independent samples of the paper-paper citation network obtained with the above procedure, which are then used to compute the average fraction of reciprocated weight for each author $\langle \rho_\mathrm{null}^{(i)} \rangle$ in the null model ensemble and excess reciprocity as per Eq. \eqref{recp}.  

Let us remark that our rewiring procedure does not change the number of citations cited and received by a paper, thus the most relevant bibliometric indicators (i.e., citation counts and the $h$-index) are preserved for each author. Hence, the rewiring preserves the scholarly reputation of each individual author, and effectively amounts to probing alternative citation patterns through which it might have been produced.

\subsection{Rich-club coefficient.}

The presence of the rich-club phenomenon in a weighted network can be detected by first ranking the authors in terms of a ``richness'' parameter $r$ \cite{opsahl2008prominence}. 
Here $r$ equals the weighted in-degree of a scientist in the author citation network, i.e. the number of received citation at a certain level of temporal aggregation. 
For each value of $r$, we select the group of all authors whose total number of received citations is larger than $r$. 
We thus obtain a series of increasingly selective clubs.
For each of such clubs, we count the number $E_{>r}$ of links connecting the members, and measure the sum $W_{>r} $ of the weights attached to such links. 
We then measure the ratio $\phi^w(r)$ between $W_{>r}$ and the sum of the weights attached to the top $E_{>r}$ strongest citations within the whole network. 
We thus measure the fraction of weights shared by the most successful authors compared with the total amount they could share if they were connected through the strongest links of the network:

\begin{equation}
	\phi_0^w(r) = \frac{W_{>r}}{\sum_{l=1}^{E_{>r}}w^{\mathrm{rank}}_l},
	\label{rich}
\end{equation}

where $w^\mathrm{rank}_l \geq w^\mathrm{rank}_{l+1}$ with $l = 1, 2, ..., E$ are the ranked weights on the citations of the network, and $E$ is the total number of links. The weighted rich-club effect can be detected by measuring the ratio

\begin{equation}
	\phi^w(r) = \frac{\phi_0^w(r)}{\phi_{\mathrm{null}}^w(r)},
\end{equation}

where $\phi_{\mathrm{null}}^w(r)$ refers to the average weighted rich-club effect assessed on the null model. 
When $\phi^w$ is larger than $1$, it means that leading authors are concentrating most of their citations towards other successful authors compared with what happens in the random null model. 
Conversely, if it is smaller than $1$, the citations among club members are weaker than randomly expected.

\section{Linear regression analysis}
\label{sec:regression}

In \ref{fig::net_stat}, we detail the results of a few linear regression models for authors whose career started between $1970$ and $1990$. The dependent variable is the average number of citations per paper for models M1-2, while the dependent variable for models M3-4 is the average total number of citations per author. In both cases those quantities are computed at the $20$th career year for each author. The average number of publications per author is used as an independent variable in all models. Moreover, the average number of citations at the $10$th career year and the average number of citations at the $10$th career year are used as independent variables for models M1-2 and M3-4, respectively. In all columns, variables such that $p < .05$ are highlighted with one asterisk, while variables such that $p < .01$ are highlighted with two asterisks. Excess reciprocity is used as an additional independent variable for models M2 and M4. As it can be seen from the $R$-squared values, the introduction of excess reciprocity as an additional independent variable does not provide any meaningful statistical improvement to the models. Yet, let us remark that in model M2, i.e., the only one where excess reciprocity is statistically significant, the relationship between excess reciprocity and success (as measured by average citations) is aligned with the results we find in the main text. 

\begin{table*}[b]
	\centering
	\caption{Linear regressions models for the relationship between citations and excess reciprocity ($1970$-$1990$).} 
	\begin{tabular}{ccccc}\\
		\hline \hline
		Variable & M1 &  M2 &  M3&  M4 \\ 
		\hline
		Intercept & ${1.97}^{**}$& ${2.31}^{**}$& ${ -55.19}^{**}$ & ${-51.55}^{**}$\\		
		Publications & ${0.02}^{**}$ & ${0.02}^{**}$ & ${3.73}^{**}$& ${3.74}^{**}$\\	
		Citations & & & ${2.37}^{**}$ & ${2.36}^{**}$  \\		
		Average Citations &  ${2.05}^{**}$ & ${2.03}^{**}$& \\	
		Excess reciprocity & & ${-1.26}^{**}$ & & $-15.11$ \\	
		\hline 
		$R$-squared & $0.5715$ & $0.5715$ & $0.6969$ & $0.6969 $\\	
		\hline \hline
	\end{tabular}
	\label{fig::net_stat}
\end{table*}	

\section{Individual excess reciprocity for $1950$-$1970$ and $1970$-$1990$}
\label{sec:ind_excess}

\begin{figure*}[h!]
	\centering
	\includegraphics[width=0.8\linewidth]{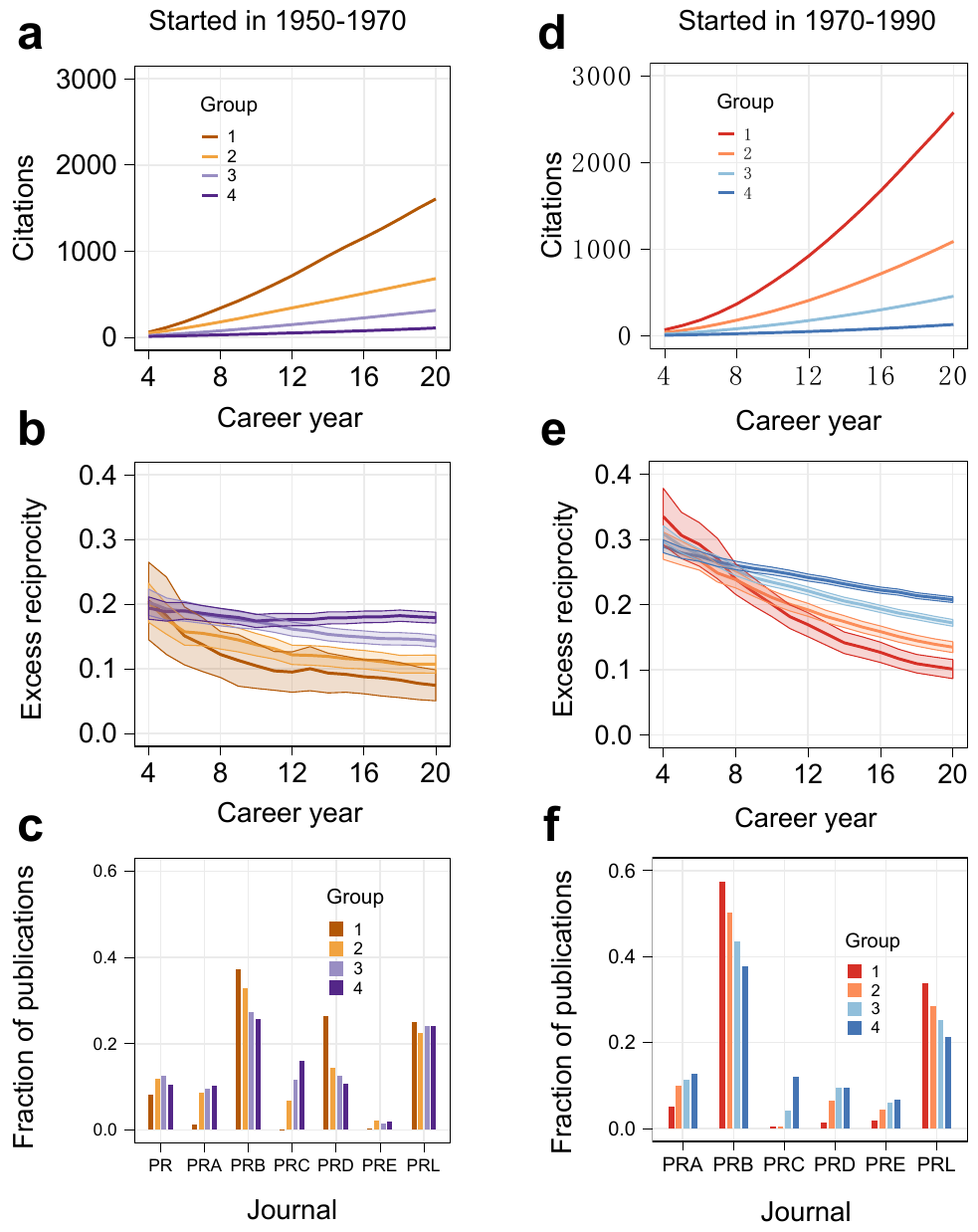}
	\caption{Illustration of the relationship between reciprocity and long-term career success ($1950$-$1970$ and $1970$-$1990$).
		\textbf{a} Centroids of the clusters identified by $k$-means based on the cumulative number of citations received by authors.  
		\textbf{b} Excess reciprocity, defined as per Eq. \eqref{recp}, within each cluster identified in \textbf{a}. Thick solid lines denote the average within the cluster, while ribbon bands denote $95\%$ confidence level intervals. 
		\textbf{c, f} Fraction of publications in each APS journal by authors from each cluster.}
	\label{fig::SM_kmeans}
\end{figure*}

We complement our analysis at the level of individual authors by considering authors who started their careers in two time periods: $1950$-$1970$ and $1970$-$1990$. In Fig. \ref{fig::SM_kmeans}, plots in the left panel are for authors who started their career between $1950$ and $1970$; Plots in the right panel are for authors who started their career between $1970$ and $1990$. The plots in panels \textbf{d} and \textbf{e} are the same shown in Fig. \ref{fig:micro} of the main text, which we replot here for convenience. We refer the reader to the caption of such Figure for the details on such plots. The sample of authors starting their career in $1950$-$1990$ contains $1,577$ authors, and the fraction of authors falling within each cluster are as follows: $3.0\%$ in group $1$, $8.2\%$ in group $2$, $28.3\%$ in group $3$, $62.1\%$ in group $4$, for authors who started their career between $1950$ and $1970$.  

\section{Alternative null model specifications}
\label{sec:alt_null}

\begin{figure*}[h!]
	\centering
	\includegraphics[width=0.9\linewidth]{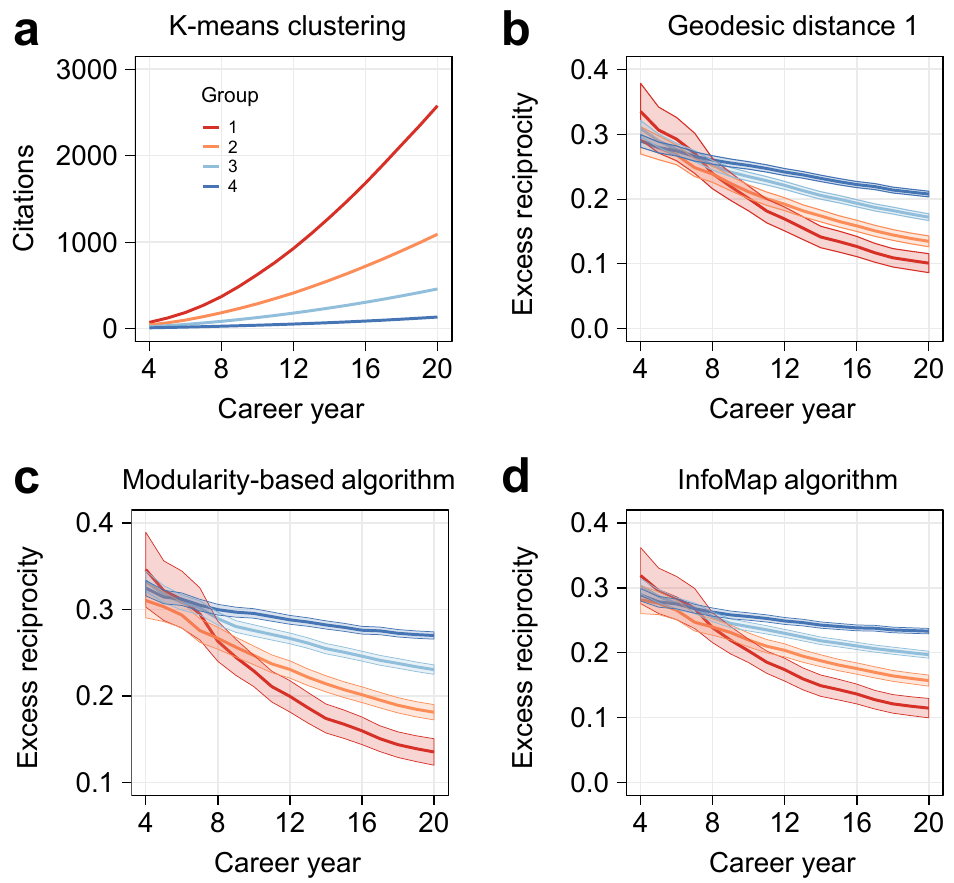}	
	\caption{Excess reciprocity for authors whose career started in $1970$-$1990$, under three different null model settings. 
	\textbf{a} Centroids of $k$-means clusters on the total number of citations from the $4$th to the $20$th author career year. 
	\textbf{b} Annual average excess reciprocity computed from the geodesic distance $1$ null model for the four clusters of authors in \textbf{a}.
	\textbf{c} Annual average excess reciprocity computed from the community--based null model with \emph{modularity-based} algorithm.
	\textbf{d} Annual average excess reciprocity computed from the community--based null model \emph{InfoMap} algorithm.
	In panels \textbf{b} and \textbf{d} thick solid lines denote the average within the cluster, while ribbon bands denote $95\%$ confidence level intervals.
	}
	\label{fig::SM_allData}
\end{figure*}

We apply a number of additional robustness checks to corroborate the excess reciprocity analysis, focusing mainly on authors that started their careers in $1970$-$1990$. 
In addition to the geodesic distance $1$ null model, we also examine two additional null models based on community structure \cite{guimera2007classes}. 
In Fig. \ref{fig::SM_allData} \textbf{c}, we show results based on the $47$ communities in the citation network of authors identified by a \emph{modularity-based} algorithm \cite{clauset2004finding}, and in Fig. \ref{fig::SM_allData} \textbf{d} we show the results obtained on the $2,458$ communities identified by the \emph{InfoMap} algorithm\cite{rosvall2008maps}. 
As described in Appendix \ref{sec:methods}, these additional null models based execute the previous rewiring procedures with constraints based on the time ordering of publications and on the community structures of author citation networks.
After the system has reached equilibrium, we compute the annual excess reciprocity for authors who published at least $10$ papers, published at least $1$ paper every five years, and whose careers lasted at least $20$ years. 
Then we use the $k$-means classification method to identify four clusters of authors based on their career citation dynamics in \textbf{a}, and see how it associates with their excess reciprocity.
The results indicate that our excess reciprocity measure is robust in all null model settings, and that excess reciprocity is negatively correlated with success at the later career stages.

\begin{figure*}[h!]
	\centering
	\includegraphics[width=0.9\linewidth]{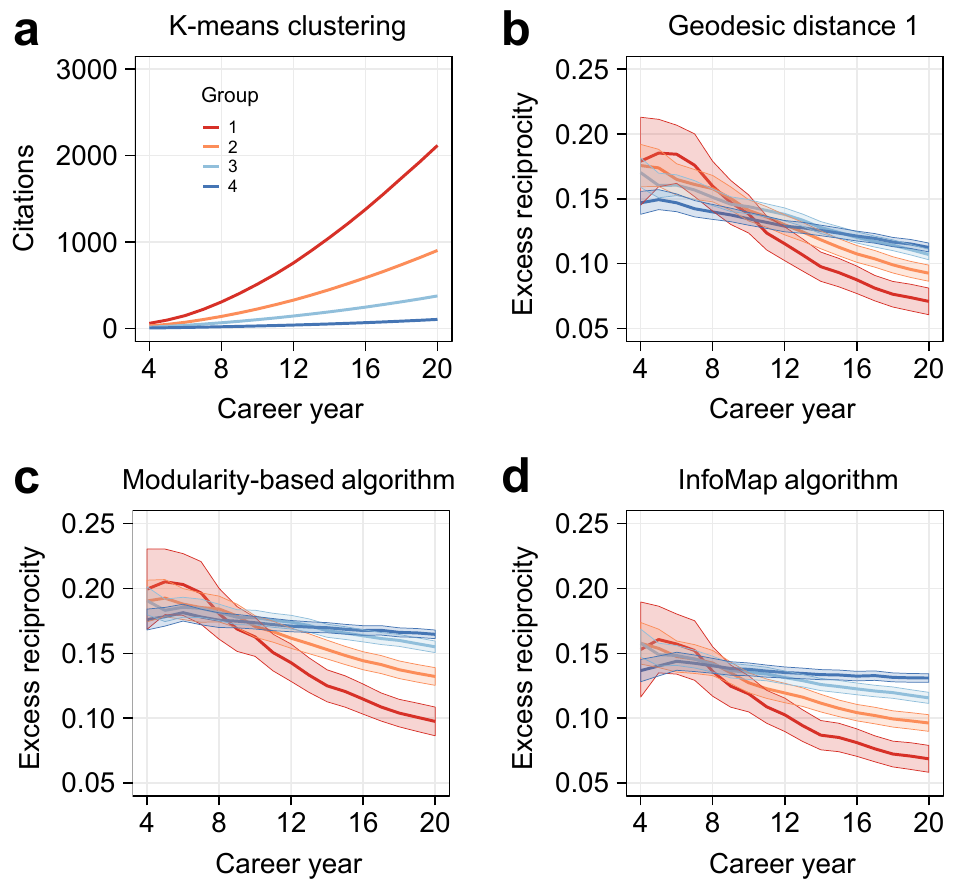}	
	\caption{Excess reciprocity for authors whose career started in $1970$-$1990$ after removing all self-citations.
		\textbf{a} Centroids of $k$-means clusters on the total number of citations from the $4$th to the $20$th author career year after removing all self-citations. 
		\textbf{a} Centroids of $k$-means clusters on the total number of citations from the $4$th to the $20$th author career year. 
		\textbf{b} Annual average excess reciprocity computed from the geodesic distance $1$ null model for the four clusters of authors in \textbf{a}.
		\textbf{c} Annual average excess reciprocity computed from the community--based null model with \emph{modularity-based} algorithm.
		\textbf{d} Annual average excess reciprocity computed from the community--based null model \emph{InfoMap} algorithm.
		In panels \textbf{b} and \textbf{d} thick solid lines denote the average within the cluster, while ribbon bands denote $95\%$ confidence level intervals.		
	}
	\label{fig::SM_noSelfCit}
\end{figure*}

We repeat this exercise by removing all self-citations from the dataset in order to examine whether our conclusions still hold when discounting self-citation as a potential driver of reciprocated citations (Fig. \ref{fig::SM_noSelfCit}).
The results show that even if the effect due to self-citations is ignored, excess reciprocity is still negatively correlated with success.

\begin{figure*}[h!]
	\centering
	\includegraphics[width=0.9\linewidth]{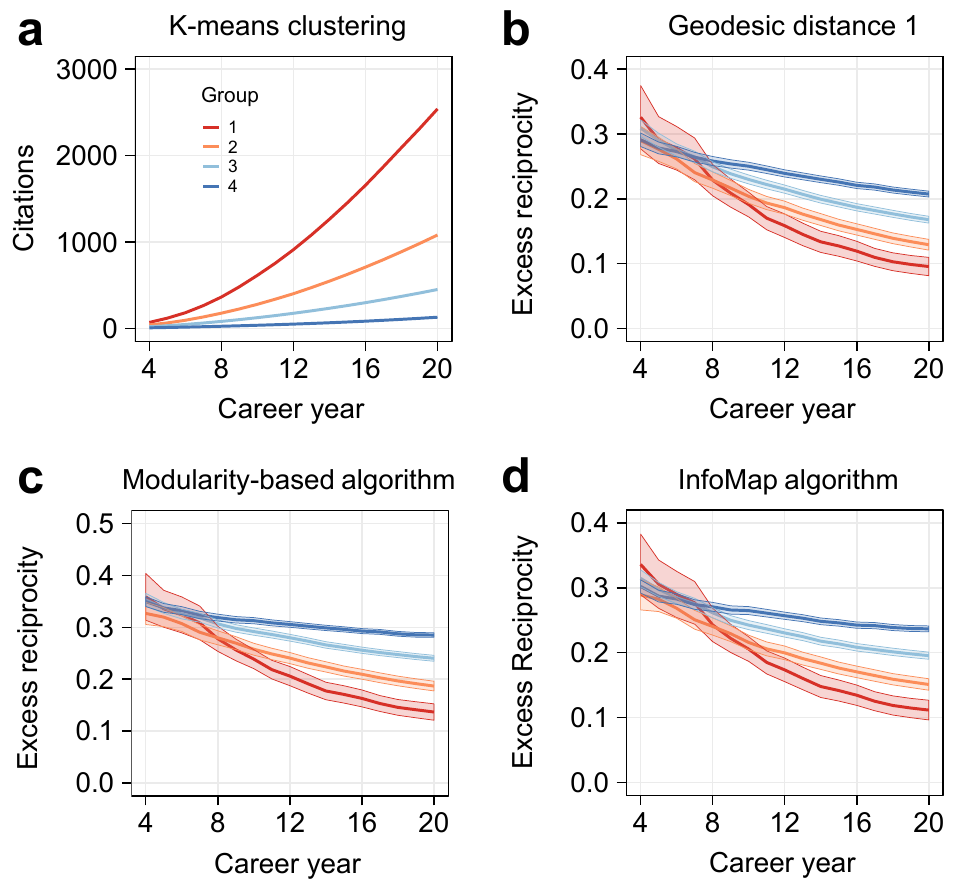}	
	\caption{Excess reciprocity for authors who started their careers in $1970$-$1990$ and have at least $10$ accrued citations by the $20$th career year. 
		\textbf{a} Centroids of $k$-means clusters on the total number of citations from the $4$th to the $20$th author career year after removing all self-citations. 
		\textbf{a} Centroids of $k$-means clusters on the total number of citations from the $4$th to the $20$th author career year. 
		\textbf{b} Annual average excess reciprocity computed from the geodesic distance $1$ null model for the four clusters of authors in \textbf{a}.
		\textbf{c} Annual average excess reciprocity computed from the community--based null model with \emph{modularity-based} algorithm.
		\textbf{d} Annual average excess reciprocity computed from the community--based null model \emph{InfoMap} algorithm.
		In panels \textbf{b} and \textbf{d} thick solid lines denote the average within the cluster, while ribbon bands denote $95\%$ confidence level intervals.		
	}
	\label{fig::SM_10citations}
\end{figure*}

We also replicate our analysis based on a productivity threshold on the authors.
In Fig. \ref{fig::SM_10citations}, we show results for null models where we select authors that have received at least $10$ citations by last career year. As most authors who have a consistently long career are able to achieve that, the sample contains $5,016$ authors, and the fractions of authors falling within each cluster are as follows: $1.3\%$ in group $1$, $7.0\%$ in group $2$, $24.9\%$ in group $3$, $66.8\%$ in group $4$.  
The negative relationship between excess reciprocity and success is still verified under this model setting.

\begin{figure*}[h!]
	\centering
	\includegraphics[width=0.9\linewidth]{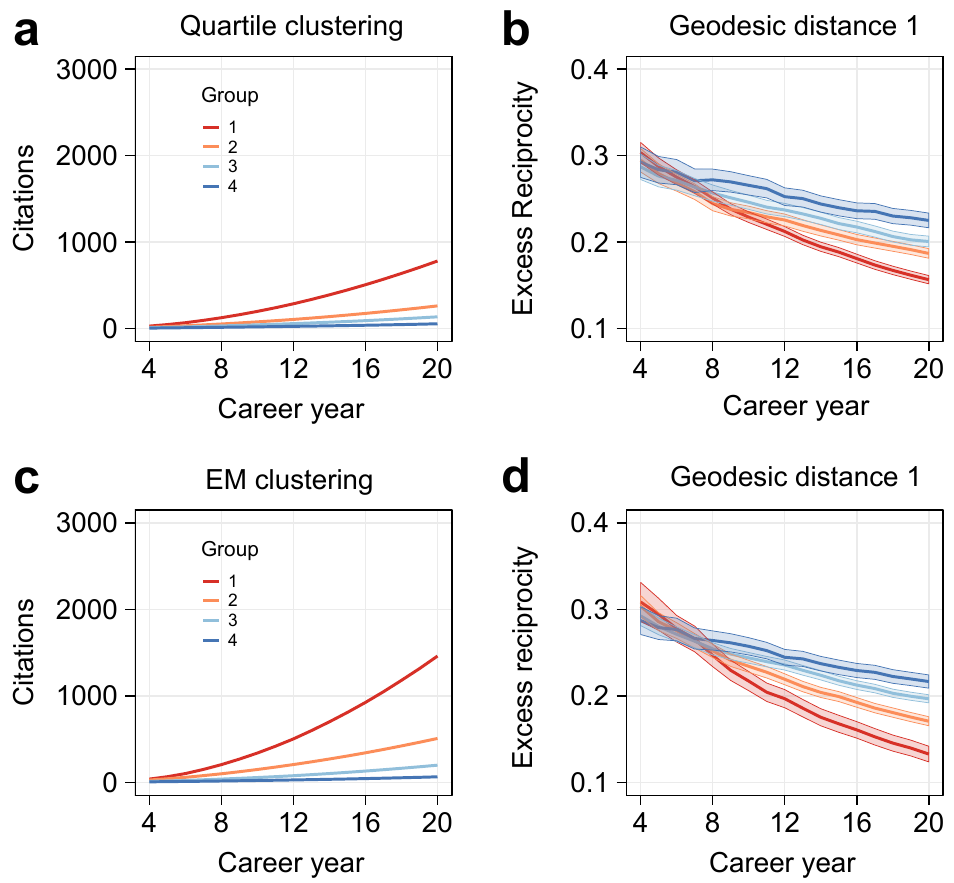}	
	\caption{Different clustering algorithms on citation patterns and excess reciprocity for authors whose career started in $1970$-$1990$.
	\textbf{a} Average career trajectories for $4$ groups of authors corresponding to the quartiles of the distribution of citations accrued over the first $20$ career years.   
	\textbf{b} Annual average excess reciprocity computed from the geodesic distance $1$ null model for the four clusters of authors in \textbf{a}.	
	\textbf{c} Average career trajectories for $4$ groups of authors obtained with the Expectation-Maximization clustering algorithm.
	\textbf{d} Annual average excess reciprocity computed from the geodesic distance $1$ null model for the four clusters of authors in \textbf{c}.
	In panels \textbf{b} and \textbf{d} thick solid lines denote the average within the cluster, while ribbon bands denote $95\%$ confidence level intervals.	
	}
	\label{fig::SM_clusteringMethods}
\end{figure*}

Finally, we study whether the clustering algorithms on author career trajectories affect on our main conclusions or not.
In Fig. \ref{fig::SM_clusteringMethods}, we cluster authors based both on the quartiles of the distribution of citations accrued over the first $20$ career years, and on the output of the Expectation-Maximization clustering algorithm  \cite{dempster1977maximum} on annual citation counts from the $4$th to the $20$th career year. 
For the the Expectation-Maximization clustering algorithm, the fractions of authors falling within each cluster are as follows: $6.0\%$ in group $1$, $23.5\%$ in group $2$, $41.8\%$ in group $3$, $28.6\%$ in group $4$.  
We find that these clustering methods do not change our conclusion that more successful careers are correlated with lower level of excess reciprocity.

\section{Matched pair analysis}
\label{sec:matched}

\begin{figure*}[h!]
	\centering
	\includegraphics[width=0.55\linewidth]{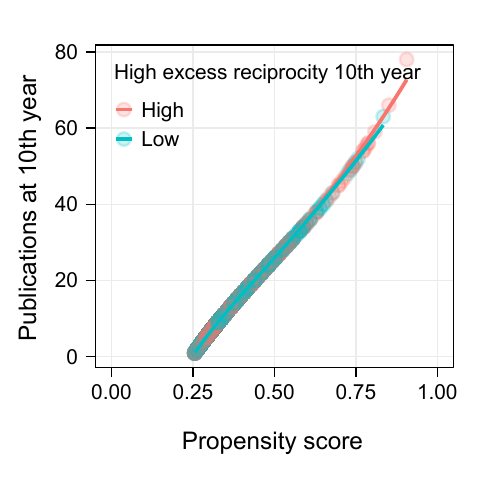}	
	\caption{Matched pair analysis of authors with high and low excess reciprocity. 	
	}
	\label{fig::SM_matchedPair}
\end{figure*}

To conduct a matched pair analysis, we first select authors whose careers started in 1970-1990 \cite{stuart2011matchit}. We separate them into two groups, i.e., we consider authors who are in the top $25\%$ in terms of excess reciprocity computed over the first $10$ career years as the treatment group, and authors in the bottom $50\%$ as the control group, which we then proceed to pair based on the number of publications in APS journals in their first $10$ career years. This leaves us with $2,536$ authors. Fig. \ref{fig::SM_matchedPair} shows the mean number of publications at the 10th career year against the propensity score (estimated running a logit model), based on the different treatment status (red for high excess reciprocity, green for low excess reciprocity). The treatment and control groups have nearly identical means at each value of the propensity score. We then proceed to measure the effects of low / high excess reciprocity on success, quantified by the number of citations received between the $11$th and $20$th career years. We find that authors in the control (low excess reciprocity) group have received, on average, $331.6$ citations, while authors in the treatment group (high excess reciprocity) have received $272.2$ citations. The difference between such two values is significant ($p < 0.001$) under a $t$-test.

\section{Decade level analyses}
\label{sec:decade}

\begin{figure*}[h!]
	\centering
	\includegraphics[width=0.55\linewidth]{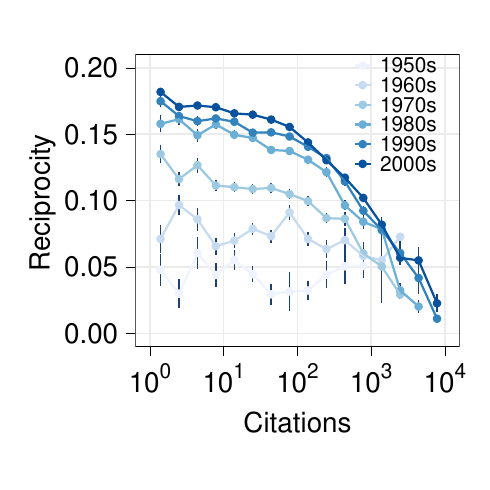}\\
	\caption{Results at the level of groups of authors for all decades.
	 	Excess reciprocity as a function of the number of citations received for authors active in each decade from the $1950$s to the $2000$s, with standard error bars.
	}
	\label{fig::recp_excess_all}
\end{figure*}

In Fig. \ref{fig::recp_excess_all}, we plot the excess reciprocity of groups of authors in six decades. It is evident that over time the overall level of excess reciprocity for low and mid-level impact authors has been steadily increasing, while the high impact group of authors did not change their behaviour substantially.

\begin{figure*}[h!]
	\centering
	\includegraphics[width=0.96\linewidth]{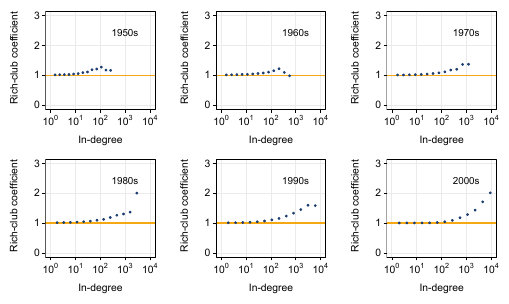}
	\caption{The rich-club effect in the empirical citation networks for all decades. Rich-club coefficient of the APS author citation network with respect to the null model (see Appendix \ref{sec:methods}) for all decades from the $1950$s to the $2000$s.
	}
	\label{fig::rich_club}
\end{figure*}

In Fig. \ref{fig::rich_club}, we plot the rich-club coefficient \cite{opsahl2008prominence} for the citation networks of authors over six decades. The rich-club effect has become more pronounced over time, suggesting that high-impact authors tend to give more citations to already successful peers.

\section{Null model constraints}
\label{sec:constraints}

\begin{figure*}[h!]
	\centering
	\includegraphics[width=0.48\linewidth]{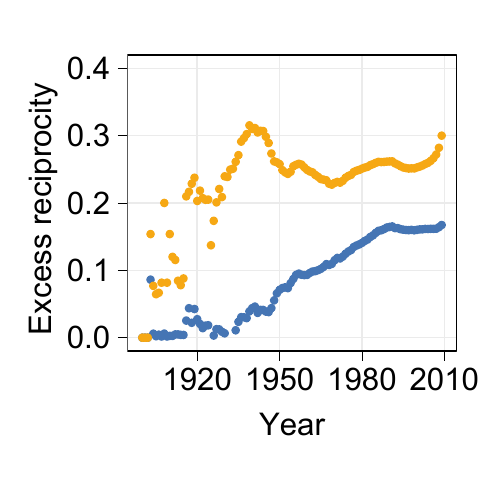}	
	\includegraphics[width=0.48\linewidth]{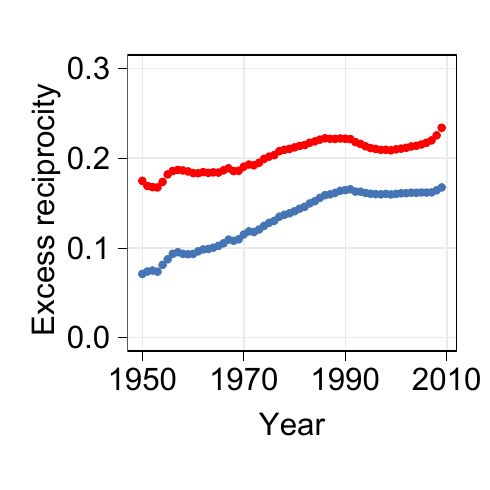}
	\caption{Additional results on network-wide excess reciprocity of the citation networks of authors.
		(Left panel): Orange dots are the reciprocity $\rho_0$ not rescaled by the null model (i.e., the fraction of reciprocated weight in the network), from 1900 to 2010.
		(Right panel): Red dots denote the excess reciprocity rescaled by a null model accounting for the time dynamics of publications but not subject to the constraint on distance implemented in the null model used in the main text.
	}
	\label{fig::recp_year_SI}
\end{figure*}

Fig. \ref{fig::recp_year_SI} shows the effect of the constraints induced by our geodesic distance $1$ null model. Blue dots in the right panel denote the the excess reciprocity $\rho$ rescaled by such null model (according to Eq. \eqref{recp} when generalized to the entire network), while red dots denote excess reciprocity when rescaled according to a null model with no constraints on the distance between authors (i.e., only with a time constraint based on the time ordering on publications, see Appendix \ref{sec:methods}). As it can be seen, the latter, less constrained, null model results in much lower levels of excess reciprocity.

\clearpage

\section*{Acknowledgements}
We thank V. Latora and D. Fernandez-Reyes for helpful comments. W.L. and G.L. were supported by an EPSRC Early Career Fellowship in Digital Economy (Grant No. EP/N006062/1).

\bibliography{scibib}

\end{document}